\documentclass[3p,onecolumn,12pt]{elsarticle}

\usepackage{amssymb}
\usepackage{mathtools, cuted}
\usepackage{hyperref}
\usepackage{xcolor}

\journal{arXiv}
\begin{document}
\begin{frontmatter}


\title{Interplay between pitch control and top speed in soccer: The stamina factor}


\author[inst1,inst2]{Marco F. da Silva}
\author[inst3]{A. Novillo}
\author[inst1,inst2]{A. Aleta}
\author[inst5]{R. López del Campo}
\author[inst5]{R. Resta}
\author[inst1,inst2]{Y. Moreno}
\author[inst3,inst4]{J. M. Buldú}

\affiliation[inst1]{organization={Institute for Biocomputation and Physics of Complex Systems (BIFI), University of Zaragoza},
            city={Zaragoza},
            country={Spain}}
\affiliation[inst2]{organization={Department of Theoretical Physics, Faculty of Sciences, University of Zaragoza},
            city={Zaragoza},
            country={Spain}}
\affiliation[inst3]{organization={Complex Systems Group, Universidad Rey Juan Carlos (URJC)},
            city={Madrid},
            country={Spain}}         
 \affiliation[inst5]{organization={Mediacoach, LaLiga},
 city={Madrid}, country={Spain}}
  \affiliation[inst4]{organization={Grupo Interdisciplinar de Sistemas Complejos (GISC)},
 city={Madrid}, country={Spain}}

\begin{abstract}
In this study, we investigate the interplay between player speed and ball control in soccer. We present a novel pitch control algorithm that quantifies the probability of a player gaining possession at any location on the field. Our model accounts for the heterogeneity of player speeds by measuring performance during matches and assigning each player a specific top speed. We then compare the pitch control percentages derived from our approach with those from classical models, which assume uniform top speeds for all players, and analyze the results across different player roles (defenders, midfielders, and forwards).
Our findings reveal a positive correlation between a player's top speed and their accumulated pitch control, with certain players benefiting more from this relationship. However, this positive correlation is constrained by the role of the player in the team, with defenders achieving the highest accumulated pitch control despite not being the fastest. Furthermore, our methodology supports team-level analysis, identifying which teams gain the greatest advantage from their players' top speeds, and extends to comparisons between the first and second halves of matches. 
Our model also enables exploration of how changes in top speed may affect pitch control at both the individual and team levels. To facilitate this, we introduce the {\it stamina factor}, a parameter that adjusts a player's top speed. We find that the impact of the stamina factor on pitch control follows a logarithmic function, with the scaling factor quantifying the potential benefits of increased speed. Interestingly, the influence of the stamina factor varies significantly by player position. Overall, our approach provides valuable insights into which teams or players could benefit most from improvements in physical performance.

\end{abstract}




\end{frontmatter}


\section{Introduction}

One of the applications of complexity sciences is the study of patterns movements of animals, herds, or humans, aiming to understand the underlying principles governing collective behaviors across species \cite{okubo1986,sumpter2006,vicsek2010}. This interdisciplinary task has evolved over decades, with significant contributions that have deepened our understanding of how collective movement arises from the particular dynamics of its agents. In 1971, W.D. Hamilton introduced the ``selfish herd theory," indicating that individuals within a group position themselves to minimize their predation risk, leading to the emergence of cohesive group formations \cite{hamilton1971}. This foundational concept highlighted how individual actions could result in complex group dynamics. Building upon this, in 1995, Vicsek et al. developed a model demonstrating how simple behavioral rules could lead to the spontaneous emergence of collective motion in groups of self-propelled particles \cite{vicsek1995}. This model described the significance of local interactions in the formation of large-scale patterns observed in animal groups. Since this seminal work, a diversity of studies, covering a variety of disciplines, focused on identifying the laws describing the essential aspects of collective motion, discussing experiments, mathematical methods, and models for simulations \cite{vicsek2010}. Interestingly, all this literature points to the fact that, despite the multidisciplinary nature of the field, similar laws describe systems of completely different nature, ranging from macromolecules to groups of animals and people. 

In recent years, the principles derived from the study of collective animal behavior have been increasingly applied to the analysis of human activities, particularly in sports. Team sports, such as soccer, exhibit complex dynamics where individual and collective movements are crucial for performance \cite{gudmundsson2017}. Understanding these dynamics through the lens of complexity science has lead to new perspectives on performance analysis and strategy development \cite{balague2013,buldu2021}. 
 This approach acknowledges the intricate interactions and emergent behaviors that arise from individual actions within the team context \cite{memmert2017}. Advancements in tracking technologies have enabled access to detailed data on player/athlete movements. Tracking datasets contain the precise Euclidean location of players during the different phases of training or competition, which, in turn, can be used to calculate their speed and acceleration. Sports such as basketball \cite{cervone2016,seidl2018}, rugby \cite{bridge2022} or soccer \cite{rahimian2022}, have benefited from the analysis of tracking datasets, extracting information about not only the physical performance of athletes and players but also the tactical implications of the individual and collective patterns of movement \cite{torresronda2022,koval2023}.
 
Furthermore, tracking datasets has facilitated analyses that go beyond traditional statistics, allowing for a deeper understanding of spatial and temporal patterns in player behavior. For instance, in \cite{novillo2024}, the location of professional soccer players during the attacking and defensive phases of the match was used to extract their corresponding speeds. The analysis of the interplay between player speed, movement angles, and distance to the ball, offered a new (and more complex) perspective on how players move strategically during different phases of the game. It was shown how the player speed is highly dependent on ball proximity, game phase (attack, defense), or player role, and how the angles of movement reveal role-specific behaviors that align with tactical responsibilities \cite{novillo2024}. In a more general framework, the application of complexity sciences in soccer highlighted the intricate balance between individual decision-making and emergent group behaviors, reinforcing the idea that player movement is not just a sum of individual actions but rather an interactive, dynamic process influenced by teammates, opponents, and contextual factors such as game situations and tactical instructions.  For example, in  \cite{marcelino2020}, Marcelino et al. analyzed the correlation of the players' speeds to show that team coordination during soccer matches is not a fixed, pre‐determined formation but rather an emergent, self‐organized process. By analyzing players’ spatiotemporal trajectories, they demonstrated that players continuously adjust their positions in response to both teammates and opponents, resulting in dynamically shifting clusters. These collective movement patterns change in response to different game phases, particularly between offensive and defensive moments  \cite{marcelino2020}. On the other hand, in 2020, network science was used to analyze tracking datasets, and the concept of {\it tracking networks} was first introduced \cite{buldu2020}. Under this framework, tracking datasets are used to quantify different kinds of interactions between players, leading to marking networks, coordination networks, or signed proximity networks. The organization of these complex networks along a match can be tracked and different network parameters can be computed to better understand the collective movements of soccer teams and players  \cite{buldu2020, chacoma2022}. 

In this paper, we focus on a specific application of the tracking datasets: the estimation of the pitch control of a soccer team. The idea of quantifying the pitch control was first introduced by Taki and Hasegawa \cite{taki1998} with the concept of {\it dominant regions}. Departing from the player location at every moment of a match, Taki and Hasegawa developed a mechanical model to simulate the movement of players and calculate the probability of controlling the ball at any future location of the pitch. In their model, the speeds and accelerations of players were included, going one step beyond the classical analysis using Voronoi partitions (i.e., disregarding the player dynamics) as a technique to evaluate the amount of the pitch controlled by each team \cite{voronoi1908,kim2004}. Therefore, not only the proximity to the ball but also the direction of the speed is crucial to determine what player will arrive before to any location of the pitch to control the ball. Some year laters, Spearman departed from the concept of {\it dominant regions} to develop a tracking-based model of passing performance \cite{spearman2017}. Players’ speeds, acceleration, reaction times, and positions on the field were integrated to obtain a Pass Success Probability Function. Combining the ball’s flight characteristics with the players’ motion capabilities, the Spearman model showed that the probability of a pass being successful is directly related to the “window” during which the intended teammate can receive the ball before an opponent can intercept. Furthermore, the predicted pass probabilities aligned well with the observed success rates from real match data. 

One significant limitation of the models presented in \cite{taki1998} and \cite{spearman2017} is the lack of heterogeneity in player speeds. Although these numerical models incorporated actual speed data from tracking systems, they assumed a uniform top speed for all players, despite evidence that top speeds vary among individuals \cite{tuo2019}. While such a simplification may be justifiable in classical pitch control models, its effect on the true dynamics of pitch control for players and teams remains unclear. To address this gap, we have implemented a pitch control model that accounts for the actual top speeds of the players under evaluation. Using tracking datasets from 100 professional soccer matches, we first calculate each player top speed for every match and then integrate these values into our numerical model, which simulates player movement to determine pitch control. This approach offers several advantages compared to the classical method. First, it provides a more accurate evaluation of pitch control for players and teams, allowing us to assess the significance of considering a uniform speed for all players. Second, it draws conclusions regarding how differences in player top speeds relate to their roles within teams, namely,  goalkeepers, defenders, midfielders, and forwards. As we will see, the impact of player top speed is not uniform across these positions. Third, our model allows to simulate scenarios in which player top speeds are increased or decreased as a result of improvements or deteriorations in fitness. To this end, we introduce the {\it stamina factor}, a control parameter that adjusts player top speed and exhibits nonlinear behavior when correlated with pitch control. In the following sections, we define the numerical model and describe the datasets used to calculate pitch control. Next, we analyze the effects of top speed on players, teams, phases of the match (attack/defense), and halves of the match (first/second). Our analysis shows the importance of incorporating player top speed into pitch control models.

\section{Methodology}

\subsection{Datasets}

 Data collection was performed using the Tracab Optical Tracking System, which employs a network of multi-camera units positioned around the stadium. This system captures the precise location of each player on the field at 25 frames per second, with a spatial resolution of up to 10 cm \cite{tracab}. Specifically, the system is based on a stereo multi-camera configuration comprising three units, each providing a resolution of $1920 \times 1080$ pixels. These units deliver a panoramic image that enables the triangulation of the location of the players and the ball. The accuracy of datasets generated by the Tracab Optical Tracking System has been validated through comparisons with GPS data in previous studies \cite{pons(2019),felipe(2019)}. Under this framework, the tracking datasets containing the positional data of players were acquired from $100$ matches of the first division of the Spanish National League (LaLiga) during the 2019/2020 season.

\subsection{The Pitch Control Model}

Pitch control at a specific location is defined as the probability that a player or team will secure possession if the ball is sent directly there. Models of pitch control simulate both ball and player dynamics by considering their current positions and speeds. This means that effective ball control depends not only on proximity but also on the direction and velocity of the players; those moving swiftly toward the ball are likely to reach it more quickly than those approaching from a different direction who must first change their course.

To develop the pitch control model, a number of computations are required at each point on the field. These involve calculating the time needed for the ball to travel from its initial position to the target location, estimating how long it would take each player to get there, and then determining the overall probability that a team will gain control once both the ball and a player have arrived. Our approach is akin to the models proposed in \cite{spearman2017,FriendsOfTracking}. 

In our model, the ball is assumed to move at a constant speed of \( v_b = 54 \) km/h. Consequently, the ball's arrival time at any location is given by 
\[
\tau_{\text{ball}} = \frac{\Delta x_b}{v_b},
\]
where \( \Delta x_b \) is the Euclidean distance from the ball's starting position to the target location. Furthermore, as an initial assumption, all players are assigned a top speed of \( v_{max,p} = 18 \) km/h, which is considered a representative estimate of the top speed achievable when contesting for ball control. It should be noted that this uniform speed assumption is standard in classical pitch control models; however, the model will later be adapted to incorporate player-specific speed variations.
To estimate a player’s expected arrival time at a given location, denoted as \( \tau_{\text{player}}(\vec{r}, t_r) \) (with \( \vec{r} \) representing the displacement vector from the player’s starting point to the destination and \( t_r \) the reaction time), we use a two-step approximation. First, a reaction time of \( t_r = 0.7 \) seconds is assumed \cite{FriendsOfTracking}; during this interval, the player is considered to continue along their current trajectory at the same speed, reaching a position \( \vec{r}_{\text{react}} \). Once the reaction time elapses, the player is modeled as running directly toward the ball at their maximum speed \( v_{max,p} \). Finally, the player's total expected arrival time is computed by combining these two phases:
\begin{equation}
    \tau_{player}(\vec{r} , t_r) = t_r + \frac{|\vec{r} - \vec{r}_{react}|}{v_{max,p}}.
    \label{exp_arr_time}
\end{equation}

After determining the arrival times for both the ball and the players, the next step is to compute the likelihood that a player gains control of the ball at the intended target location. To achieve this, we adopt Spearman’s assumption \cite{spearman2017} that ball control is a stochastic process governed by an exponential distribution with a constant rate \(\lambda\), which is the reciprocal of the mean time needed for a player to secure possession. Consequently, during any small time interval \(\Delta t\) in which a player is near the ball, the chance of controlling it is \(\lambda \cdot \Delta t\). Consistent with \cite{spearman2017}, we set \(\lambda = 4.3\, s^{-1}\) as our baseline control rate. Moreover, goalkeepers are assigned a higher control rate, \(\lambda_{GK} = 12.9\, s^{-1}\), to reflect their enhanced ability to claim the ball, especially given their capacity to use their hands.

To further refine the model, an uncertainty parameter \(\sigma\) is introduced into the players’ arrival times to account for factors not explicitly modeled. Combining these elements, the probability \(F_{\text{int},j}(\vec{r}, t, \sigma, t_r)\) that player \(j\) intercepts the ball at time \(t\) and at location \(\vec{r}\) is described by the cumulative distribution function of a sigmoid:

\begin{equation}
F_{\text{int},j}(\vec{r}, t, \sigma, t_r) = \frac{1}{1 + e^{-\frac{t - \tau_{j}(\vec{r}; t_r)}{\sqrt{3}\sigma/\pi}}}.
\end{equation}

Note that this probability requires incorporating the actions or performance of other players on the field. Therefore, the overall probability that player \(j\) controls the ball when moving toward the location specified by the displacement vector \(\vec{r}\), in the presence of other \(k\) players, is described by the following differential equation:

\begin{equation}
\frac{dPC_j}{dt}\left(t, \vec{r}, \sigma, \lambda_j, t_r\right) = \left(1 - \sum_k PC_k\left(t, \vec{r}, \sigma, \lambda_k\right)\right) F_{\text{int},j}(t, \vec{r}, \sigma, t_r) \lambda_j,
\label{final_eq}
\end{equation}

where \(PC_j\) denotes the potential Pitch Control field for player \(j\) and \(\lambda_j\) is the individual control rate. The term \(\sum_k PC_k\left(t, \vec{r}, \sigma, \lambda_k\right)\) represents the cumulative pitch control field of all other players \(k\) on the pitch at time \(t\).

By integrating Eq. \(\ref{final_eq}\) over the interval \(t \in [t_{\text{ball}}, t_{\text{ball}} + 10]\) seconds and initializing \(PC_j\left(t, \vec{r}, \sigma, \lambda_j\right) = 0\) at the start, we obtain the control probability for each player at every point on the pitch. These values are then aggregated for all teammates to derive the overall pitch control for each team.

\subsection{Speed dependence and the {\it stamina factor}}

Conventional pitch control models assign a uniform top speed \( v_{max,p} \) to all players \cite{taki1998,spearman2017}, thereby simplifying computations at the expense of accuracy. To address this limitation, we first computed the probability distribution of each player \(i\)’s speed during every half of the match and extracted the 95th percentile threshold, denoted as \(v_{95}(i)\). We then replaced the generic parameter \(v_{max,p}\) with these player-specific values \(v_{95}(i)\). This adjustment incorporates individual running capabilities into the model, allowing each player's pitch control to reflect their reported top speed. Note that, since pitch control is effectively a ``zero-sum game", an increase in one player's control necessarily results in a decrease in another's. We call the speed-dependent pitch control model as $PC_v$.

Moreover, the capacity to run faster does not imply that a player will always utilize their top speed. Depending on the match context or their physical condition, players may choose to conserve energy rather than run at full speed. Conversely, over-motivated players may increase their top speed during specific matches. This observation prompted us to explore the effects on pitch control when a player runs at a different pace than usual, either faster or slower. To capture this variability, we introduce a  ``stamina factor” \(\xi\) into the model, which multiplies the player's top speed (i.e.,  \( v_{max,p}(i) = \xi  \cdot v_{95}(i)\)), increasing it when \(\xi > 1\) and decreasing it when \(\xi < 1\).

\section{Results}

\begin{figure}[!ht]
    \centering
    \includegraphics[width = 1.0\textwidth]{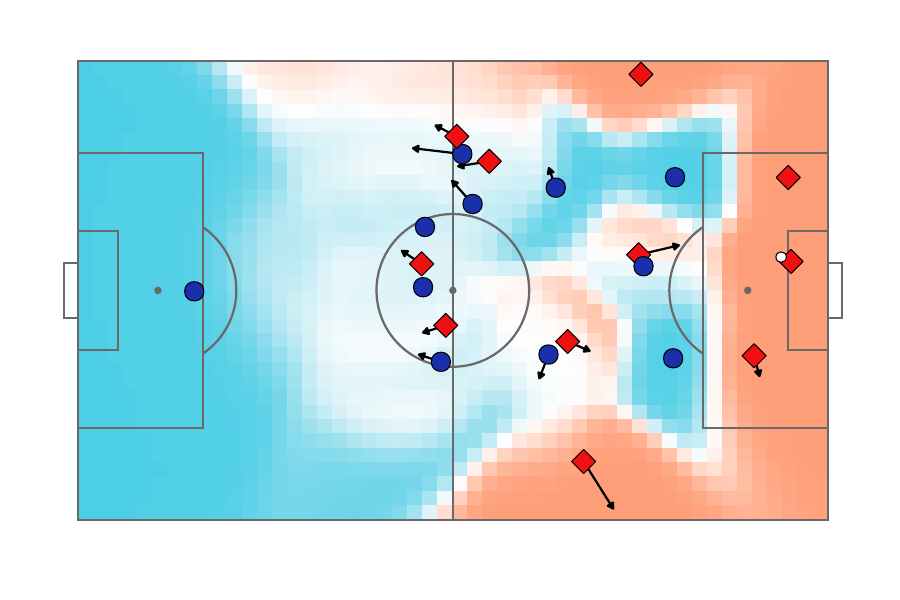}
    \caption{Example of the pitch control (PC) calculated in a given frame of a match. Colors (blue/red) indicate the team with a higher probability of controlling the ball in case it moves to any region of the pitch. The length of the arrows assigned to each player is proportional to the speed at this specific frame. The player shape (circles or squares) indicates the team they belong to.}
    \label{f01}
\end{figure}

Figure \ref{f01} shows an example of the classical pitch control calculated over a particular frame of a match. Colors (blue/red) indicate the team that would control the ball in case it reaches any region of the pitch and color intensity is proportional to the pitch control of each team. In the figure, we have included a vector indicating the direction of the speed of players, with a length proportional to the modulus. Players without an arrow are stationary. We can observe how the direction of the speed increases the probability of controlling the ball in the direction of the arrow. To elaborate this figure, the player speed is the one recorded at the specific frame of the figure, however, to simulate the ability of each player to move to other regions of the pitch to control the ball, all players are considered to run at the same speed, as it is assumed in classical pitch models. In the following sections, we will analyze the ability of players to move faster/slower than their counterparts.

\subsection{Player pitch control vs player speed}

\begin{figure}[!ht]
    \centering
    \includegraphics[width = 0.9\textwidth]{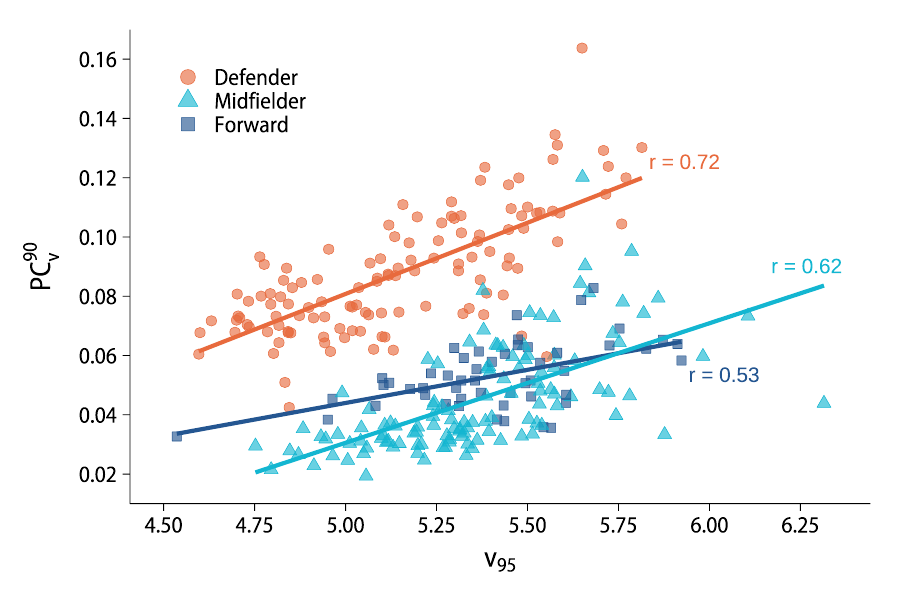}
    \caption{Normalized speed-dependent pitch control $PC^{90}_v$ as a function of the player top speed $v_{95}$. Each point corresponds to the pitch control accumulated by a player during $90$ minutes. Three player positions are considered: defenders (circles), midfielders (triangles), and forwards (squares). Note the positive correlation $r$ (Pearson correlation coefficient) in all positions.}
    \label{f02}
\end{figure}

As mentioned in previous sections, the fact that players' top speeds change from player to player has unavoidable consequences on their corresponding pitch control. Faster players are able to reach regions of the pitch than slower players when departing from similar distances. This fact is accounted for when considering the player's top speed $v_{95}$ in a speed-dependent pitch control model $PC_v$. In Fig. \ref{f02} we show the percentage of pitch control accumulated by players in 90 effective minutes of play when their observed top speed is considered. In the figure, we have classified players into 3 different positions (defenders, midfielders, and forwards) and omitted the goalkeepers since their pitch control has different implications than the rest of the players. We can observe how the pitch control of players is positively correlated with their speed for all player positions (see correlations coefficients $r$ at the figure). Interestingly, defenders, who are the players accumulating more pitch control, are the ones with the highest correlation coefficient. However, also note that they are not necessarily the players with the highest top speeds, indicating that not only the speed but also the role of the player, affects the values of the pitch control.

Figure \ref{f02} also highlights the fact that pitch control depends on more factors beyond the player top speed or the player position, as indicated by the low to moderate values of the correlation coefficients and the heterogeneity of the accumulated pitch control of players with similar top speeds.

\subsection{Team's pitch control}

\begin{figure}[!ht]
    \centering
    \includegraphics[width = 0.9\textwidth]{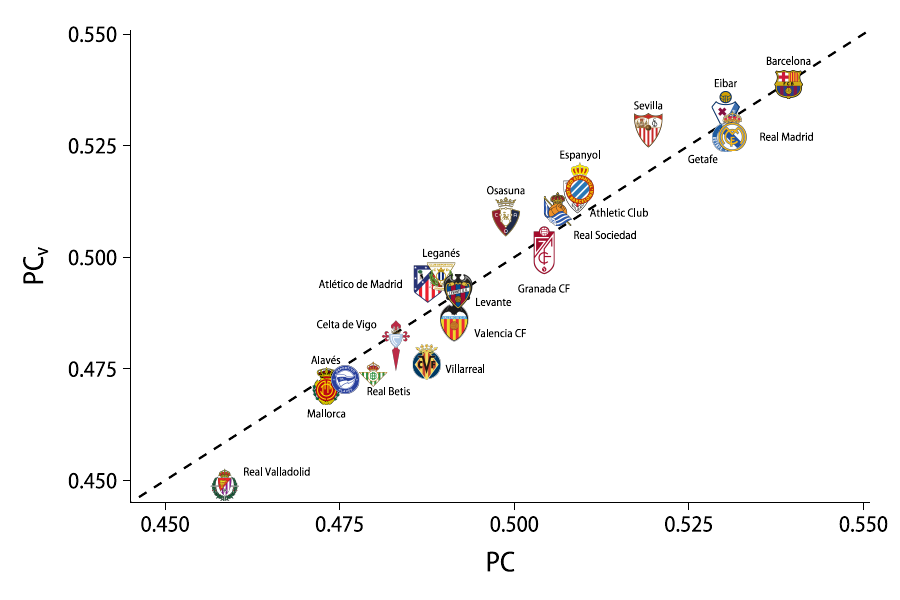}
    \caption{Classical pitch control ($PC$) vs speed-dependent pitch control ($PC_{v}$) for the Spanish teams at first division (season 2019/2020). Despite the positive correlation, there are deviations from $PC(i)=PC_v(i)$, which is indicated by the dashed line.}
    \label{f03}
\end{figure}

The consequences of including the player's top speed in the pitch control model can be projected into the team level. The fact that pitch control is a zero-sum game, where the increase of a given team results in the decrease of its rival, makes the analysis at the team level a useful tool to evaluate what teams are suffering the most from the physical performance of their players. With this aim, we analyzed how the pitch control of teams in the Spanish first division was affected by the inclusion of player top speed in the model. The pitch control of a team is defined as the sum of all team players' pitch control. Figure \ref{f03} shows a comparison of the classical pitch control ($PC$) with the speed-dependent pitch control ($PC_v$) for each single team. Both variables are averaged over $10$ matches. Note that teams with values over $0.5$ accumulate more pitch control than their rivals. 

Teams located above the identity line ($PC_v(i) > PC(i)$) see their pitch control enhanced when player speeds are taken into account, while teams below the line experience a reduction in pitch control under the new model. The interpretation of the increase/decrease of the pitch control is not straightforward. It may reflect a playing style that emphasizes tactical organization or positional discipline over high-speed transitions, or simply differences in the physical profiles of the squad. It is important to note that pitch control is only one of many aspects that contribute to team performance on the field. It is worth noting how teams like Sevilla show both high pitch control and further improvements when speed is considered, suggesting a strong alignment between their tactical execution and physical capabilities. Notably, F.C. Barcelona ranks highest in both classical and speed-dependent pitch control, highlighting the consistency of their spatial dominance regardless of the modeling approach.

\begin{figure}[!ht]
    \centering
    \includegraphics[width = 0.9\textwidth]{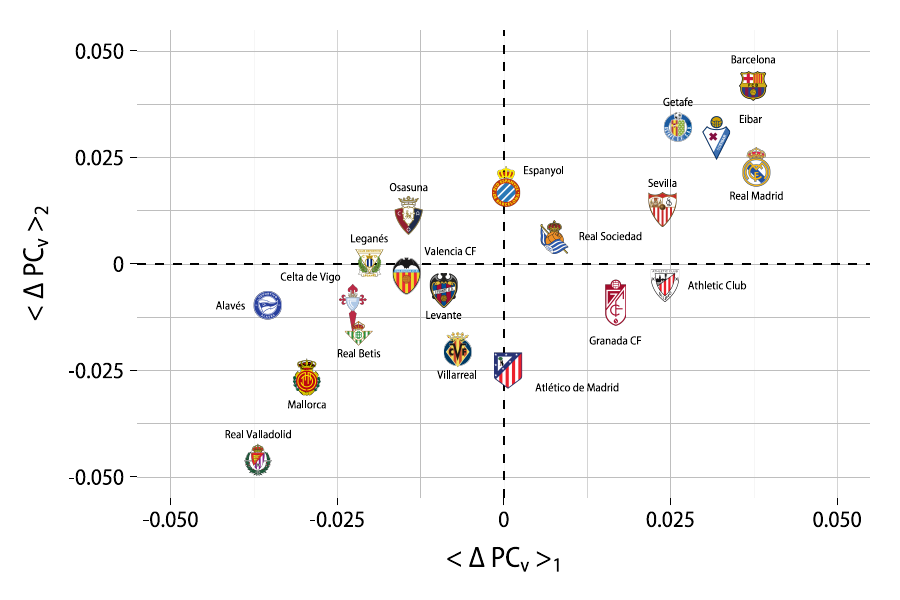}
    \caption{Pitch control of all teams in the competition depending on the part of the match. The pitch control accumulated during the first and second halves, $\langle \Delta PC_v \rangle_1$ and $\langle \Delta PC_v \rangle_2$, respectively, has been normalized by subtracting the average,  $\overline{PC}_{vx}$ of all teams at each half $x$. Dashed lines correspond to the averages of all teams at each half of the match.}
    \label{f04}
\end{figure}

The fact that pitch control can be calculated at every frame allows to investigate its evolution. For example, in Fig. \ref{f04} we have split the analysis into the two halves of a match. For each team, we normalized the pitch control at each half of a match by subtracting the average of all teams $\Delta PC_{vx} = PC_{vx} - \overline{PC}_{vx}$, where $x$ denotes the half. Then, we calculate the average over all matches of a team, $\langle \Delta PC_v\rangle_{1,2}$. This way, we are able to analyze what teams accumulated more pitch control than the others at each half of the matches.  Figure \ref{f04} shows that, despite the positive correlation between the pitch control accumulated at both halves, not all teams behave in the same way. Comparing Fig. \ref{f03} and Fig. \ref{f04} we can observe how, for example, Getafe and Real Madrid have similar pitch controls; however, Getafe generates more pitch control in the second half than Real Madrid, whose pitch control is higher, on average, in the first half of a match. Another interesting case is the one of Atl\'etico de Madrid, whose pitch control is average in the first half, but decays drastically in the second half. 

\subsection{Stamina factor}

\begin{figure}[!ht]
    \centering
    \includegraphics[width = 1.0\textwidth]{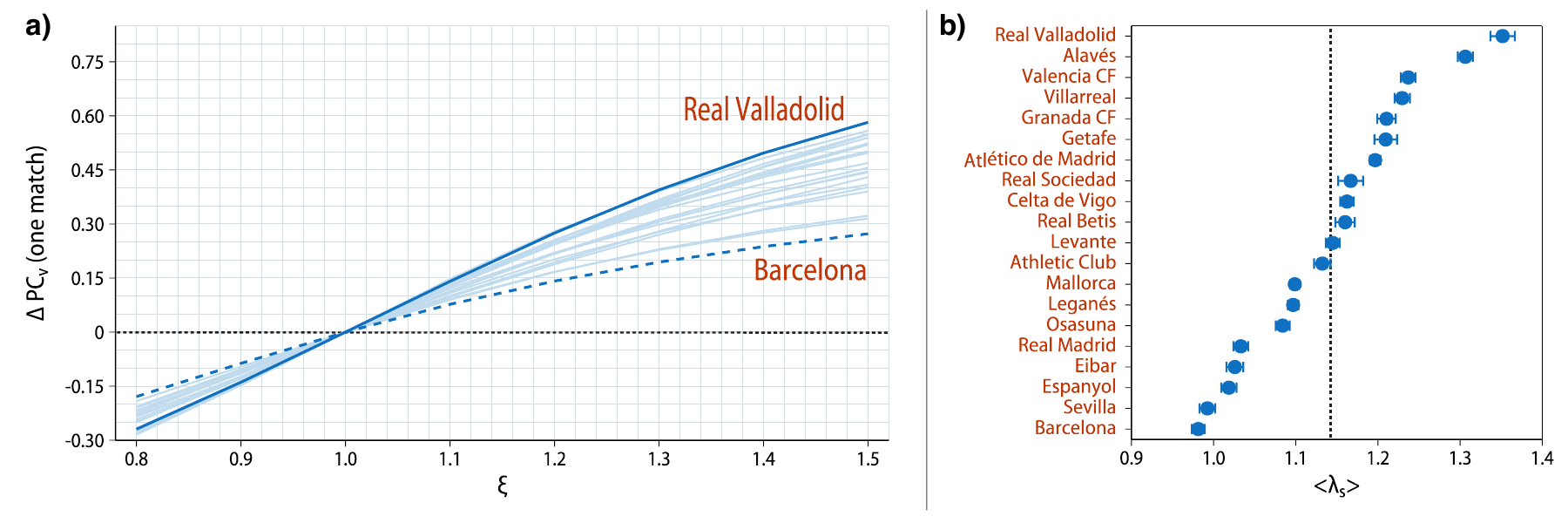}
    \caption{In (a), pitch control variation in a single match, $\Delta PC_v$, as a function of the stamina factor $\xi$, for each team. Each line corresponds to one of the twenty teams playing in the competition. We can observe how they follow a logarithmic distribution in all cases. In b), we show the average scaling factor $\langle \lambda_s \rangle$ of each team, ranked according to its value. The distribution indicates what teams would benefit the most from the stamina factor (i.e., Valencia CF, Alavés, and Real Valladolid). The vertical dashed line is the average of the scaling factor for the whole competition.}
    \label{f05}
\end{figure}

The next step is to evaluate what is the influence of increasing/decreasing player speed in the accumulated pitch control of teams. The stamina factor $\xi$ modifies the top speed of each player and the pitch control can be recalculated with the modified speeds. This way, we can analyze how pitch control would be increased/decreased when a player can run faster/slower, thus evaluating the tentative impact of improving/worsening its fitness. Figure \ref{f05}(a) shows the variation of pitch control $\Delta PC_v$ for different values of the stamina factor for one match of each team. We can observe how all teams have similar behavior, which basically consists of a positive correlation that fits with a logarithmic function of the form $f(x) = \lambda_s \log x + b$. Then, we fit this function for the $\Delta PC_v$ for each game and obtain the average scaling parameter $\langle \lambda_s \rangle$ for each team. Note that the scaling parameter $\lambda_s$ controls the dependence of the pitch control of a team on the stamina factor. Figure \ref{f05}(b) shows the ranking of teams according to its corresponding $\langle \lambda_s \rangle$. As we will discuss later, teams with a lower $\langle \lambda_s \rangle$ are the less affected by modifications of the player top speed, while teams with the highest values are those which would be benefited the most if the speed of its players was increased.  However, they would also suffer more a decrease on top speed.

\begin{figure}[!ht]
    \centering
    \includegraphics[width = 0.75\textwidth]{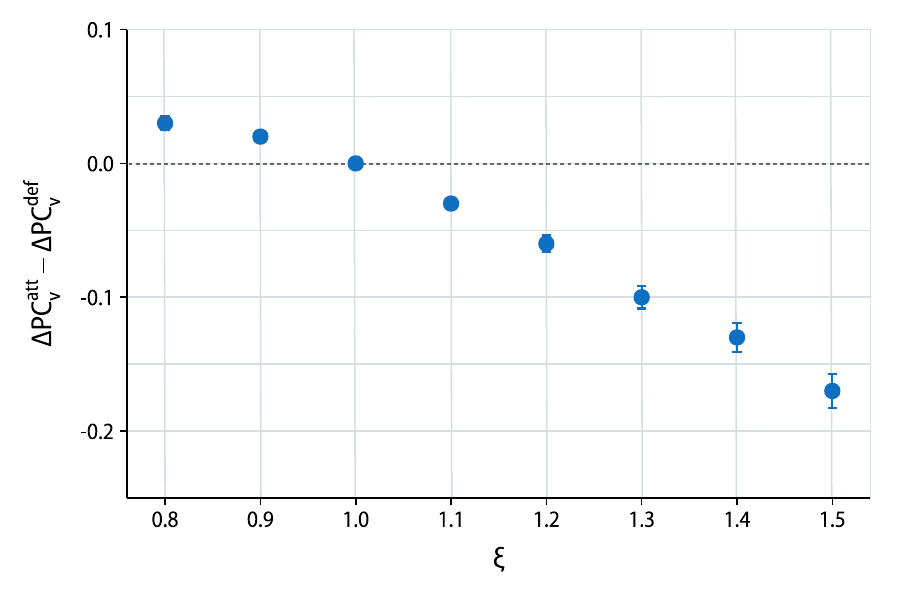}
    \caption{Effect of the stamina factor during each match phase: attack vs. defense. Difference between the increment of pitch control when attacking ($\Delta PC^{att}_v$) and defending ($\Delta PC^{def}_v$), for multiple values of the stamina factor $\xi$. Points represent the average for each team in the competition, with error bars corresponding to one standard deviation.}
    \label{f06}
\end{figure}

Furthermore, different kinds of analysis can be performed by filtering the states of the match when the increase of pitch control is calculated. For example, in Fig. \ref{f06} we distinguish the impact of the stamina factor between the attacking state and and defending state of a team. Interestingly, while the impact of the stamina factor is positive in both cases, i.e., increases pitch control for $\xi>1$ and decreases it when $\xi<1$ (not shown), the amount of the impact is not the same. The subtraction $\Delta PC_v^{att}$ - $\Delta PC_v^{def}$ yields negative values for $\xi >1$ and positive values for $\xi <1$ indicating that $\Delta PC_v^{def}$ is consistently larger in magnitude than that in $\Delta PC_v^{att}$. Note that for $\xi < 1$ both are negative, but $\Delta PC_v^{def}$ is more negative, indicating a major loss. Similarly, for $\xi > 1$ both are positive, but  $\Delta PC_v^{def}$ is larger, indicating a larger benefit.

\begin{figure}[!ht]
    \centering
    \includegraphics[width = 0.9\textwidth]{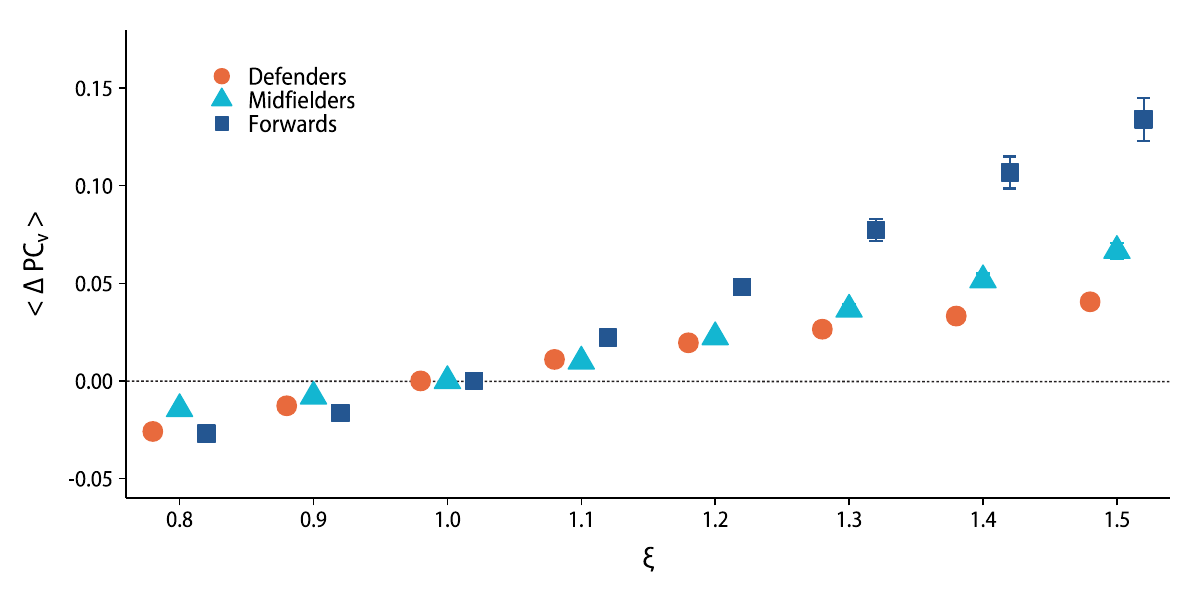}
    \caption{Team average speed-dependent pitch control variation $\langle\Delta PC_v\rangle$ vs the stamina factor $\xi$, for each role of a player. For each $\xi$ and pair team/game, we only modify the top speed of the players with the given role. Then, we estimate the variation of the pitch control, $\Delta PC_v$, and normalize it by the number of players per modified position in that team and game. Lastly, we average the variations over all games played by a team for each role considered: defenders (circles), midfielders (triangles) and forwards (squares)}
    \label{f07}
\end{figure}

Finally, it is also possible to translate the analysis of the stamina factor to the level of players. In Fig. \ref{f07} we plot the increase of pitch control as a function of the stamina factor for the three main roles of the players: defenders, midfielders and forwards. We can observe how the forward players are the most affected by the positive increments of the stamina factor, followed by midfielders and defenders. However, there exists a certain degree of asymmetry in the positive/negative increments of the stamina factor. While the forward players are the ones that, on average, would benefit the most from an increase in their speed, a decrease produces much more moderate differences. Two conclusions can be extracted from this fact. First, the effect of increasing/decreasing the top speed of players is not linear and, second, the role of the player in the team, probably constraining the location on the pitch, is strongly related with the effects of the stamina factor on the pitch control.

\clearpage
\section{Discussion}

Integrating pitch control models into soccer performance analysis leads to a significant advancement in understanding both individual player dynamics and team strategies. These models offer a more detailed perspective on how players influence the available space on the field and shape offensive and defensive actions. However, player speed is a key determinant of a team's capacity to create or restrict space, thereby influencing strategic decisions and match outcomes. In that sense, we have shown that, by accounting for variations in player top speed, pitch control models yield more precise insights into a player's contribution to collective movement and spatial dominance. Incorporating each player's specific top speed reveals differences in individual performance with regard to the pitch control accumulated over a match. Interestingly, these differences are influenced by the player's role on the team; for example, defenders tend to accumulate higher pitch control (excluding goalkeepers), even though they are not necessarily the fastest players on the field. This observation underscores the complex interplay between individual top speed and pitch control. At the collective level, our analysis demonstrates that incorporating the actual top speed data recorded during matches affects the overall pitch control accumulated by soccer teams. While there is a positive correlation between the classical pitch control and the one that contains the player top speed, deviations from this trend emerge. For example, some teams, such as Sevilla F.C., Osasuna, and Atlético de Madrid, exhibit increased pitch control when player-specific top speeds are incorporated, whereas teams like Real Valladolid or Villarreal experience a decrease. This fact is not good or bad by itself, but indicates that not considering the player top-speed leads to an inaccurate quantification of pitch-control that is specially significant in certain teams. Our results point to the importance of using actual top speed and suggest that more sophisticated models that include a ``player avatar" (i.e., containing information of each player such as the top speed, acceleration or strength) could better characterize individual and team performance. It is also noteworthy that the impact of top speed on pitch control is not uniform throughout a match. This is reported when examining the effect of player top speed on pitch control separately for each half. Our analysis reveals that not all teams consistently increase or decrease their pitch control across both halves, indicating that this phenomenon is complex and phase-dependent.  For instance, while FC Barcelona demonstrates above-average pitch control in both halves, Osasuna shows increased control in the second half. In the case of Granada, the highest pitch control is in the first half.

The observation that individual top speed influences pitch control opens the possibility of exploring how voluntary changes in top speed, whether due to motivational factors or improved physical conditioning, might affect pitch control. We observed that pitch control exhibits a logarithmic dependence on the stamina factor, a parameter that modulates player top speed. All teams increased (decreased) their accumulated pitch control for stamina factor values greater (less) than one. However, although the relationship is logarithmic for all teams, the scaling factor varies, indicating that some teams are more sensitive to changes in player top speed than others. Notably, teams such as Real Valladolid, Alavés, and Valencia CF would benefit the most from higher top speeds, whereas FC Barcelona, Sevilla, and Espanyol would experience smaller gains.

Another finding that highlights the complexity of the analysis is the different impact of the stamina factor during defensive versus offensive phases. Our results indicate that for positive values of \(\xi\), the increase in pitch control is more pronounced during defensive phases than during attacking ones. This suggests that enhancing players top speed may be particularly beneficial for defensive efforts, a consideration that could be taken into account when programming the player physical loads during training. Moreover, the impact of the stamina factor appears to be closely tied to a player's role on the team. Forwards benefit the most from an increase in top speed, while defenders experience the smallest improvements in pitch control. This may be attributed to the distinct positional roles and movement patterns on the field: defenders typically maintain a steady pace, whereas forwards are more likely to engage in intermittent bursts of acceleration to outmaneuver opponents.

In light of these findings, we believe that analyzing player speeds in relation to pitch control is an intriguing area with applications that extend beyond the scope of this paper. Future work could involve developing comprehensive player avatars that incorporate detailed information on physical and tactical performance, technical abilities, and, where feasible, psychological attributes. Such an approach would enable a better understanding of the complex behavior of soccer players and, more broadly, any team sport where spatiotemporal interactions between teammates and opponents are crucial. As soccer continues to evolve toward data-driven analysis, pitch control models that consider the complexity of the problem and contain the player specific kinematics will be essential for advancing performance evaluation, tactical decision-making, and team coordination.

\section{Acknowledgments}

J.M.B. acknowledges the support by Ministerio de Ciencia e Innovaci\'on (Spain) under grants PID2020-113737GB-I00 and PID2023-147827NB-I00. A.A. acknowledges support from the Grant No. RYC2021-033226-I funded by MCIN/AEI/10.13039/501100011033 and the European Union NextGenerationEU/PRTR. Y.M was partially supported by the Government of Aragón, Spain, and “ERDF A way of making Europe” through Grant No. E36-23R (FENOL). A.A. and Y.M. acknowledge the support by Ministerio de Ciencia e Innovación, Agencia Española de Investigación (MCIN/AEI/10.13039/501100011033) Grant No. PID2023-149409NB-I00. M.F. acknowledges the support by the Government of Aragón through a Government of Aragón Ph.D. contract order ECU/592/2024.

\section{Code availability}

The pitch control algorithm utilized to generate these results is accessible at:
\url{https://github.com/Markfds01/Tracking_laliga.git}. Due to confidentiality agreements, the dataset used in this study cannot be publicly disclosed.



\end{document}